\documentclass[fleqn,10pt]{wlscirep}
\usepackage{lineno}
\usepackage{multirow}
\usepackage{enumitem}
\usepackage{cases}
\usepackage{color}
\newcommand{\el}{El Ni\~{n}o}

\title{
Complexity based approach for \el ~magnitude forecasting before the ``spring predictability barrier'' 
}

\author[1+]{Jun Meng}
\author[1+*]{Jingfang Fan}
\author[1+]{Josef Ludescher}
\author[1]{Agarwal Ankit}
\author[2]{Xiaosong Chen}
\author[3]{Armin Bunde}
\author[1,4]{Jürgen Kurths}
\author[1*]{Hans Joachim Schellnhuber}
\affil[1]{Potsdam Institute for Climate Impact Research, 14412 Potsdam, Germany}
\affil[2]{School of Systems Science, Beijing Normal University, 100875 Beijing, China}
\affil[3]{Institut f\"{u}r Theoretische Physik, Justus-Liebig-Universit\"{a}t Giessen, 35392 Giessen, Germany}
\affil[4]{Department of Physics, Humboldt University, 10099 Berlin, Germany}

\affil[+]{these authors contributed equally to this work}

\affil[*]{jingfang@pik-potsdam.de or john@pik-potsdam.de}

\begin{abstract}
The \el ~Southern Oscillation (ENSO) is one of the most prominent interannual climate phenomena. An early and reliable ENSO forecasting remains a crucial goal, due to its serious implications for economy, society, and ecosystem. Despite the development of various dynamical and statistical prediction models in the recent decades, the ``spring predictability barrier'' (SPB) remains a great challenge for long (over 6-month) lead-time forecasting. To overcome this barrier, here we develop an analysis tool, the System Sample Entropy (SysSampEn), to measure the complexity (disorder) of the system composed of temperature anomaly time series in the Ni\~{n}o 3.4 region. When applying this tool to several near surface air-temperature and sea surface temperature datasets, we find that in all datasets a strong positive correlation exists between the magnitude of \el ~and the previous calendar year's SysSampEn (complexity). We show that this correlation allows to forecast the magnitude of an \el ~with a prediction horizon of 1 year and high accuracy (i.e., Root Mean Square Error $=0.23^\circ C$ for the average of the individual datasets forecasts). For the on-going 2018 \el ~event, our method forecasts a weak \el ~with a magnitude of $1.11\pm 0.23^\circ C$.
Our framework presented here not only facilitates a long--term forecasting of the \el ~magnitude but can potentially also be used as a measure for the complexity of other natural or engineering complex systems.

\end{abstract}
\begin{document}

\flushbottom
\maketitle
%
%
\thispagestyle{empty}
%

\section*{Introduction}

\begin{figure}
\begin{centering}
\includegraphics[width=0.9\linewidth]{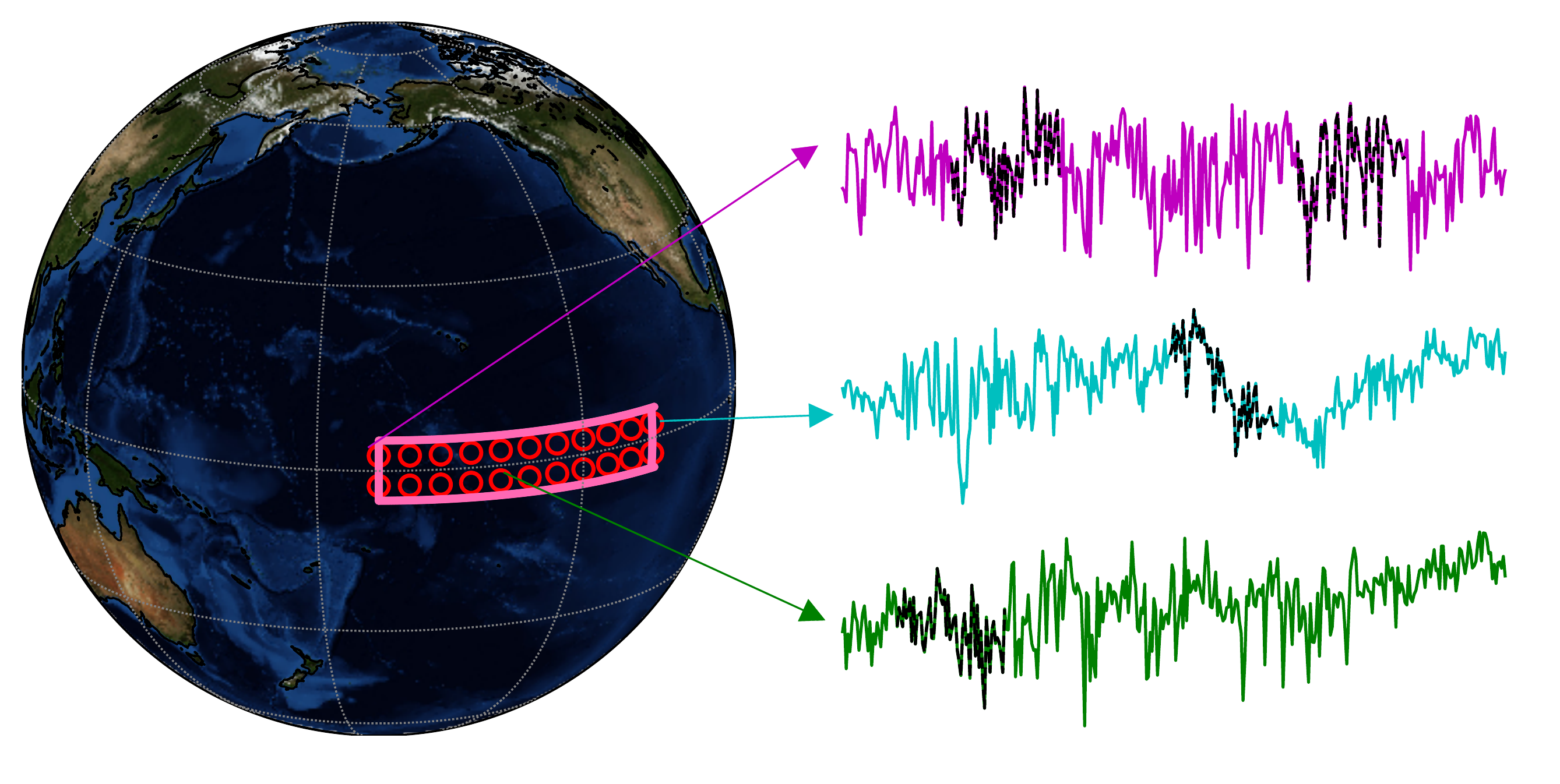}
\caption{\label{Fig1}The Ni\~{n}o 3.4 region. The red circles indicate the $22$ nodes in the Ni\~{n}o 3.4 region with a spatial resolution of $5^\circ \times 5^\circ$. The curves are examples of the temperature anomaly time series for three nodes in the Ni\~{n}o 3.4 region for one specific year, and several examples of their sub-sequences are marked in black.
 }
\par\end{centering}
\end{figure}

ENSO, the interannual fluctuation between anomalous warm and cold conditions in the tropical Pacific, is one of the most influential coupled ocean-atmosphere climate phenomena on Earth \cite{Dijkstra2005,McPhaden2006,Clarke2008,Cane2010}. The warm phase of ENSO (\el) is characterized by an abnormal warming of the eastern equatorial Pacific, which occurs about every 2-7 years. The Oceanic Niño Index \cite{ONI} (ONI) is the primary indicator that the  National Oceanic and Atmospheric Administration (NOAA) uses to monitor and identify ENSO events. It is the 3-month running mean of sea surface temperature (SST) anomalies in the Ni\~{n}o 3.4 region ($5^\circ S -5^\circ N$, $170^\circ W - 120^\circ W$, shown in Fig. \ref{Fig1} as the region inside the pink rectangle). An \el ~event is defined to take place if the ONI is at or above $0.5^\circ C$ for at least $5$ consecutive months (red shades in Fig. \ref{Fig2}a). Here we use the value of the highest peak of the ONI during an \el ~event, to quantify its magnitude.

\el ~has been reported to affect the marine ecosystems, commercial fisheries, agriculture, public safety, and to even bring extreme weather conditions in many parts of the globe \cite{Ropelewski1987,Kiladis1989,Ropelewski1992,Diaz2001,Kumar2006,Hsiang2011,Burke2015,Schleussner2016,Fan2017
}. Thus the understanding of the underlying mechanism and prediction of \el ~are of great importance for humanity. Numerous models, dynamical as well as statistical ones, were developed to simulate and forecast \el ~events. Dynamical models \cite{Zebiak1987,McCreary1991,Kleeman1993,Kleeman1995,Wang1996,Jin1997a,Jin1997b,Wang1999
,Palmer2004,Saha2006} express mathematically the physical equations of the ocean--atmosphere system. In contrast, statistical model \cite{Xu1990,Penland1993} forecasts of \el ~are based on data-driven analyses. During the past decades, the prediction of \el ~has made great progress and skillful forecasts at shorter lead times (up to around 6 months) are possible \cite{Kirtman2002,Chen2008,kurths2019}. However, both types of models reveal very low predictability before and during boreal spring (February-May). This is the so-called ``spring predictability barrier'' (SPB) \cite{Webster1992,Lau1996,McPhaden2003,McPhaden2012}.

Recently, several approaches based on climate networks were developed to forecast the onsets of \el ~around one year in advance \cite{josef2013,jun2017,jun2018,Nooteboom2018}. One of these approaches\cite{josef2013} has correctly forecasted all \el ~onsets or their absence since 2012. 
However, this method is unable to predict the magnitude of the event. Predicting the magnitude is crucial since a stronger \el ~usually causes more extreme events, e.g., floods, droughts or severe storms, which have serious consequences for economies, societies, and ecosystems. In particular, the \el ~events which started in 1997 and 2014 exhibited relatively high magnitudes and had major impacts on the dynamics and structure of the tropical and temperate ecosystems worldwide \cite{hughes2017coral}. To fill this gap, here we develop an analysis tool, the \textit{System Sample Entropy (SysSampEn)}, to quantify the spatio-temporal disorder degree of temperature variations in the Ni\~{n}o 3.4 region, and to forecast the \el ~magnitude before the SPB. 
Based on a calendar year's data we forecast, if in the following year an \el ~will start or not. Once the SysSampEn approach forecasted the occurrence of an \el ~onset, we are able to forecast its magnitude with high skill (i.e., correlation $r=0.84$ and $RMSE=0.23^\circ C$ between the forecasted and observed magnitudes for the \el ~events that occurred during the last 35 years). 
We like to mention that the SysSampEn approach roughly doubles the lead-time at comparable skill. The skill of our \el ~magnitude forecast, based on the previous year's SysSampEn, and thus with a lead-time of about 1 year, is comparable to the best state-of-the-art model forecasts which start in June (i.e., with 6-month lead-time) and predict the same year's boreal winter (November-January) ONI \cite{Barnston2012,Wang2018}.

\section*{System Sample Entropy}
We define the SysSampEn for a complex system as a generalization of Sample entropy (SampEn) \cite{Richman2000} and Cross-SampEn \cite{Richman2000}.
SampEn was introduced as a modification of approximate entropy \cite{Pincus1991,Pincus1995a}. It measures the complexity related to the Kolmogorov entropy \cite{Kolmogorov1958}, the rate of information production, of a process represented by single time series. The Cross-SampEn was introduced to measure the degree of asynchrony or dissimilarity between two related time series \cite{Richman2000,Pincus1995b}. Both have been widely used in physiological fields, e.g., to make early diagnoses before the clinical signs of neonatal sepsis by analyzing the heart rate variability \cite{Lake}, to implement an automatic diagnosis of epileptic EEG \cite{Acharya}, and to discriminate different sensory conditions by analyzing human postural sway data \cite{Ramdani}.

However, a complex system such as the climate system is usually composed of several related time series (e.g., curves in Fig. \ref{Fig1}).
Therefore, here we introduce the SysSampEn as a measure of the system complexity, to quantify simultaneously the mean temporal disorder degree of all the time series in a complex system, as well as the asynchrony among them. Specifically, it approximately equals to the negative natural logarithm of the conditional probability that two sub-sequences similar (within a certain tolerance range) for $m$ consecutive data points remain similar for the next $p$ points, where the sub-sequences can originate from either the same or different time series (e.g., black curves in Fig. \ref{Fig1}), i.e.,
\begin{equation}
SysSampEn(m,p,l_{eff},\gamma)=-log(\frac{A}{B}),
\label{eq1}
\end{equation}	
where $A$ is the number of pairs of similar sub-sequences of length $m+p$, and $B$ is the number of pairs of similar sub-sequences of length $m$, $l_{eff} \leq l$ is the number of data points used in the calculation for each time series of length $l$, and $\gamma$ is a constant which determines the tolerance range. The detailed definition of SysSampEn for an arbitrary complex system composed of $N$ time series is described in the Method section. When $N=1$, $p=1$, and $l_{eff}=l$, our definition is equivalent to the classical SampEn \cite{Richman2000}.
As it is the case for SampEn and Cross-SampEn, before the SysSampEn can be used as an effective tool, 
appropriate parameter values have to be identified since only certain value combinations can be used to estimate a system's complexity with considerable accuracy. The effective parameter combinations may be different in different complex systems. Here we choose $m$ to be 30 days or 60 days and $p$ to be 15 days or 30 days since \el ~is an interannual phenomenon. 

\section*{Results}

\subsection*{Strong positive correlation between the \el ~magnitude and its previous calendar year's SysSampEn}

\begin{figure}
\begin{centering}
\includegraphics[width=0.9\linewidth]{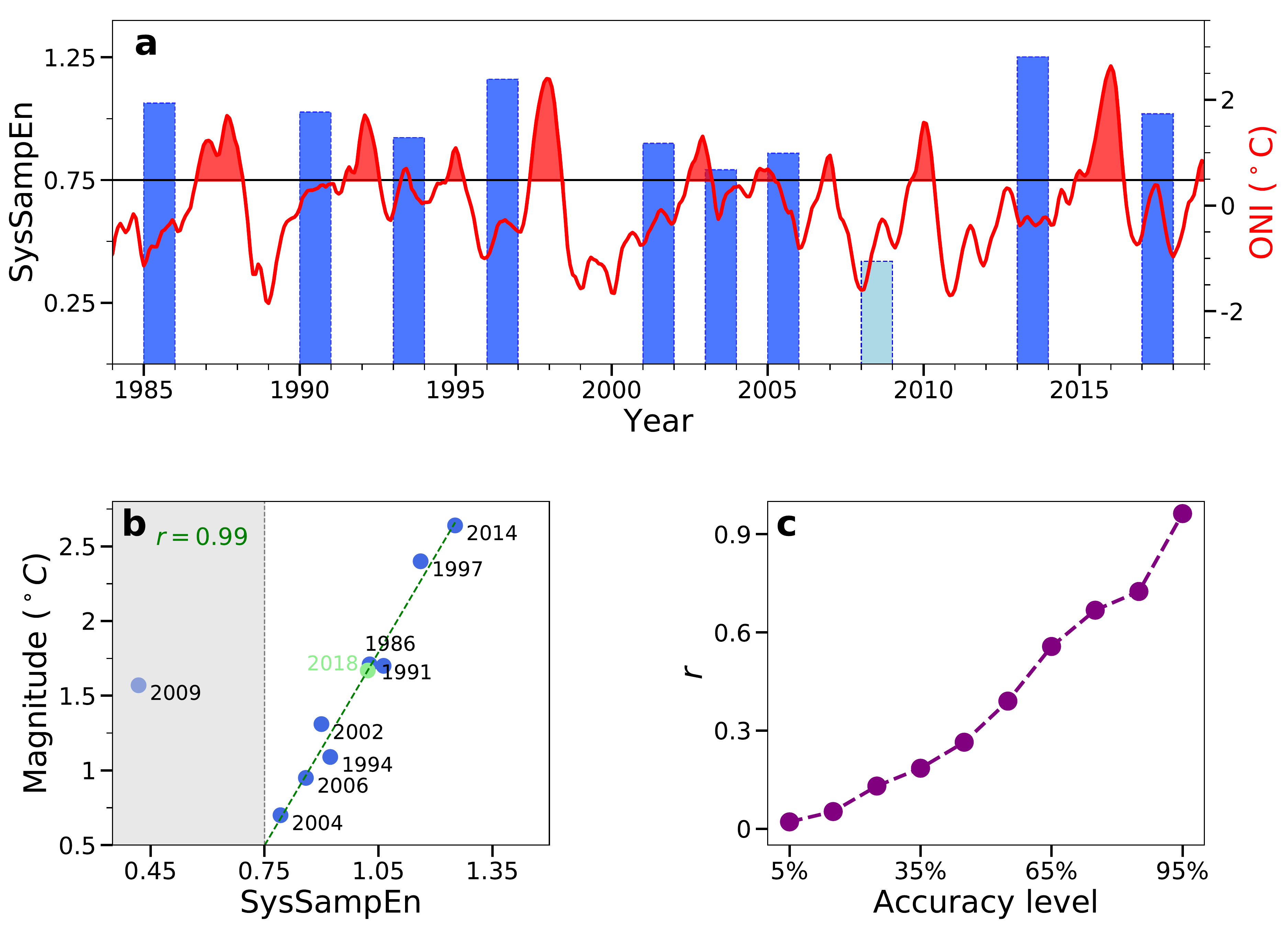}
\caption{\label{Fig2}Correlation between SysSampEn and \el ~magnitude. \textbf{a} The heights of the blue rectangles indicate the values of the SysSampEn (left scale) for the calendar years preceding \el ~events, calculated from ERA-Interim, by using the set of parameters ($m=60\ days$, $p=15\ day$s, $l_{eff}=345\ days$ and $\gamma=9$) that correspond to the highest correlation $r$ with \el ~magnitudes. The red curve is the ONI and the red shades indicate \el ~periods (right scale). \textbf{b} Scatter plot of the maximal \el ~magnitude versus previous calendar year's SysSampEn (blue rectangles in \textbf{a}). The gray region indicates values of the SysSampEn, which predict for the maximal ONI less than $0.5^\circ C$ and thus by definition non-\el ~events. The green dashed line shows the best least-square fitted line. \textbf{c} The y-coordinate of each purple dot is the averaged correlation $r$ for parameter combinations with accuracy no less than a certain level (,i.e., its x-coordinate), in both the spatial asynchrony and the temporal disorder tests. 
The correlation $r$ between SysSampEn and the \el ~magnitude is monotonously increasing with increasing accuracy level.
The calculation of the accuracy level is independent of any \el ~events, thus the strong correlation between the SysSampEn and the \el ~magnitudes emerges naturally without fitting.}
\par\end{centering}
\end{figure}

We calculate the SysSampEn of the climate system composed of the near surface air or sea surface temperature anomaly time series in the Ni\~{n}o 3.4 region and find a strong positive correlation between the \el ~magnitude and the SysSampEn of its previous calendar year (Fig. \ref{Fig2}a,b). 
This positive correlation is significant ($r=0.90$ on average) and robust across all the analysed datasets (ERA-Interim 1000hPa air temperature \cite{ERA-Interim} (ERA-Interim), ERA5 1000hPa air temparature \cite{ERA5} (ERA5), ERA5 sea surface temperature (ERA5 SST) and JRA55-do sea surface temperature \cite{JRA-do} (JRA55-do SST)) (Fig. S1). 

In the following, we present our results based on the dataset of ERA-Interim, which gives the highest correlation. For a given calendar year between 1984 and 2018, we construct a system composed of temperature anomaly time series in the Ni\~{n}o 3.4 region (see Fig.~\ref{Fig1}) with a spatial resolution of $5^\circ \times 5^\circ$.

First, we determine the parameter combinations for the SysSampEn, which enable an accurate estimation of the system's complexity.  
We do this by performing two tests (for details see Method section), which determine, for a given parameter combination, the ability of the SysSampEn to discriminate between higher and lower disordered systems. In the temporal disorder test, we add random numbers to the real temperature data, while in spatial asynchrony test, we compare two systems, one which is constructed from neighboring points on the globe and one which is constructed from randomly chosen points on the globe. An accurate complexity measure should be able to recognize the higher disorder in the more random system and thus assign a higher SysSampEn value to it. We define accuracy as the percentage of correct assignments. Thus, using suitable parameter combinations for the SysSampEn we can quantify the temporal, as well as the spatial disorder in the system.

Surprisingly, we find that the previous calendar year's SysSampEn exhibits a strong positive correlation with the magnitude of \el, if the parameter combination for the SysSampEn can quantify the system complexity with good accuracy. 
Fig. \ref{Fig2}c demonstrates on one example, $m=60\ days$ and $p=15\ days$, that with changing the values of $l_{eff}$ and $\gamma$ in Eq. \ref{eq1}, the Pearson correlation ($r$) between the \el ~magnitude and the previous calendar year's SysSampEn (e.g., blue rectangles in Fig. \ref{Fig2}a) increases significantly with the accuracy level. Please note, that the accuracies are calculated fully independently of any \el ~magnitude analyses or forecasts. Thus, the strong correlation between the SysSampEn and the \el ~magnitude is naturally obtained from the parameter combinations, which enable the SysSampEn to quantify the system complexity with high accuracy. In other words, the high predictability of the \el ~magnitude before the “spring predictability barrier” is not the result of overfitting, but it originates from the strong and robust correlation between system complexity and \el ~magnitude.

We also find that the pattern of the SysSampEn between 1984 and 2018 is independent of the data resolution and highly consistent for different parameter combinations which provide high accuracy (Fig. S2 and Table. S1). In particular, the correlation between the \el ~magnitude and the previous calendar year's SysSampEn with different effective ($\geq 95\%$ accuracy level) parameters are all significantly high (the average $r$ is $0.83\pm0.12$), while the best correlation $r=0.99$ is obtained for $m=60\ days$, $p=15\ days$, $l_{eff}=345\ days$ and $\gamma=9$ (Fig. \ref{Fig2}b).

We perform the same analysis on the other datasets and obtain similar results, see Figs. S3-S5 in the SI. We also present in Fig. S1 the scatter plots of the \el ~magnitude versus the previous calendar year's SysSampEn that give the highest $r$ for each of the other three datasets. The correlation $r$ is also significantly high for the other three datasets, and the average $r$ when using all high accuracy parameter combinations of the four datasets (Tables. S1) is $0.79\pm0.11$. Note that the 2009 \el ~is the only event missed in the onset forecasts (discussed below) and is an exception in the linear relationship.

To obtain the best forecasting performance, we choose the SysSampEn parameters by first conducting an accuracy test and only accepting parameter combinations which lead to a high accuracy (accuracy level $\geq 95\%$ for air temperature and $\geq 85\%$ for SST) in both the spatial asynchrony and the temporal disorder tests. From these high accuracy parameter combinations, we choose in the second step, the one which gives the highest correlation $r$ with the magnitudes of the past \el ~events. We repeat this for all datasets. Table \ref{table} shows the parameters that suggest the highest $r$ for \el ~events before 2018 in the different datasets.

We like to note that, using the old entropy definitions to quantify the system complexity, i.e., calculating the average SampEn per node or the average Cross-SampEn for each pair of nodes in the Nino 3.4 region, we get less significant correlations ($r = 0.42$ on average) than in the SysSampEn approach.


\subsection*{Forecasts of \el ~magnitudes and onsets}

Based on the substantial correlations between SysSampEn and \el ~magnitude, we develop efficient hindcasting and forecasting methods for both the \el ~onsets and magnitudes (introduced in the Method section). Then we forecast the magnitude of the ongoing 2018 event, by utilizing the previous calendar year's temperatures.

\begin{figure}
\begin{centering}
\includegraphics[width=0.9\linewidth]{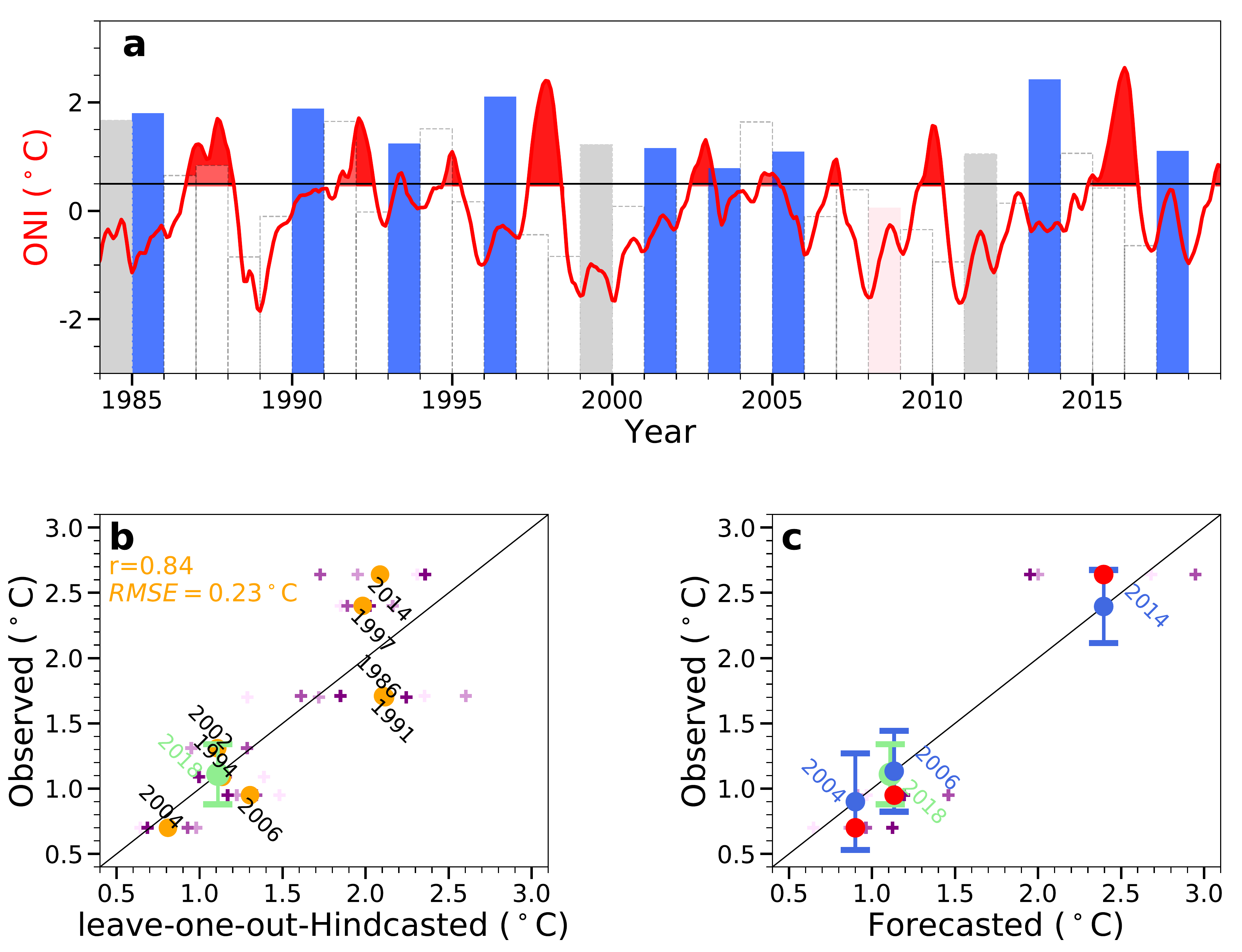}
\caption{\label{Fig3} Forecasting the \el ~onsets and magnitudes \textbf{a} 
The value of onset forecasting index (average forecast over the four datasets) is shown as the height of rectangles and is used to forecast the occurrence or absence of an \el ~onset in the following year. 
If the index value is $\geq0.5^\circ C$ and the observed ONI in December is below $0.5^\circ C$, we forecast the onset of an \el ~in the following year. 
The blue rectangles show the correctly forecasted \el ~onsets, the pink rectangle indicates a missed \el ~event, the gray rectangles indicate false alarms and the transparent rectangles show when the absence of an \el ~onset was correctly forecasted. 
\textbf{b} 
Observed temperature versus the leave-one-out hindcasted temperature for the \el ~magnitudes (orange dots) before 2018. The obtained RMSE is $0.23^\circ C$. The forecasted magnitude ($1.11^\circ C$) of the 2018 \el ~event is plotted as a light green dot with an error bar of $1\times RMSE$.  The ``plus'' symbols indicate the hindcasted (forecasted) values obtained by using each of the four datasets. \textbf{c} Forecasts of the 2004, 2006 and 2014 El Ni\~{n}o magnitudes based only on past information. The error bar for each forecasted \el ~event (blue points) equals $1\times RMSE$ (i.e., $0.37^\circ C$, $0.31^\circ C$ and $0.28^\circ C$ for the 2004, 2006 and 2014 events, respectively) and is calculated from the leave-one-out hindcasts which lie in the regarded events past. Thus, the forecasted value, as well as its error bar (i.e., $1\times RMSE$), are only based on the event's past information. The red dots show the observed magnitudes and are within the error bars. The forecasted 2018 magnitude and its error bar are shown in light green.}
\par\end{centering}
\end{figure}

To show the high predictability of the \el ~onset before the SPB, we compose a new index (rectangles in Fig. \ref{Fig3}a) by substituting the value of the SysSampEn for each calendar year into the best fitting linear functions (green dashed lines in Fig. \ref{Fig2}b and Fig. S1), and then taking the average over all the four datasets. Thus the new index has the unit of $^\circ C$. We find that the value of this index for one specific calendar year can be used to forecast the presence or absence of an \el ~onset in the following year with very good accuracy, i.e., 9 out of 10 correct forecasts of \el ~onsets (dark blue rectangles), with only one missed (pink rectangle); 21 out of 24 correct forecasts of \el ~onset absence years (transparent rectangles), with three missed (gray rectangles). The detailed algorithm is introduced in the Method section.

To demonstrate the high predictability of the \el ~magnitudes before the SPB, we firstly perform leave-one-out hindcasts (described in the Method section) of the magnitudes for all the \el ~events between 1984 and 2017. For each dataset, we use the parameter combination in the function of SysSampEn that gives the highest correlation $r$ between SysSampEn and the magnitudes of the \el ~events before 2018 (Table \ref{table}). The observed \el ~magnitudes and hindcasted magnitudes are shown in Fig. \ref{Fig3}b. Compared to the real data, we find that our hindcasting method is quite efficient with considerable accuracy, i.e., the root of mean square error (RMSE) $=0.23^\circ C$. This indicates that the SysSampEn method has the potential for skillful \el ~magnitudes forecasts with a prediction horizon of $1$ year.

Secondly, we perform magnitude forecasts for the 2004, 2006 and 2014 \el ~events by using only data from the event's past (see Method section), and find that the differences between the observed and forecasted values are within $1\times RMSE$, see Fig. \ref{Fig3}c. These results indicate that $1\times RMSE$ can be regarded as an error bar. The $RMSE$ is obtained by leave-one-out hindcasting applied only to the regarded events past, e.g., for the 2004 \el ~it depends only on the period 1984-2003. Analogously, the SysSampEn parameters also depend only on the regarded events past and are given in Tables S2-S4. Please note that, for later \el ~events, as more data becomes available for our method, the estimated RMSEs become smaller, see Fig. \ref{Fig3}c.
The forecast performance for the last 3 \el ~events demonstrates the ability of our method to forecast the \el ~magnitude, as well as providing correct error estimates.


Next, we apply the SysSampEn method to forecast the magnitude of the on-going 2018 \el ~event, based only on data up to 2017. The used SysSampEn parameters are given in Table \ref{table} and obtain for its magnitude $1.11^\circ C$, with an error bar of $0.23 ^\circ C$.

\begin{table*}[htbp]
\center
\scalebox{1.0}{
\begin{tabular}{c|c|c|c|c|c|c|c|c}
\hline
\multicolumn{3}{c|}{DATA}&\multicolumn{4}{c|}{\multirow{2}{*}{}parameter}&\multirow{2}{*}{$r$}&\multirow{2}{*}{2018($^\circ$C)}\\
\cline{1-7}
Type&Name&Resolution&$m(days)$&$p (=q)(days)$& $\gamma$&$l_{eff}(days)$&&\\
\hline
\multirow{2}{1.5cm}{T 1000hPa}&ERA-Interim&5$^\circ$&60&15&9&345&0.99&1.67\\
\cline{2-2}\cline{3-9}
&ERA5&5$^\circ$&30&30&8&330&0.87&0.58\\ 
\hline
\multirow{2}{*}{SST}&ERA5 &5$^\circ$&30&30&5&330&0.86&1.09\\
\cline{2-9}
&JRA-do&4$^\circ$&30&30&5&360&0.87&1.09\\
\hline
\multicolumn{7}{c|}{\textbf{Average}}&\textbf{0.90}&\textbf{1.11}\\
\hline
\end{tabular}
}
\caption{\label{table} Values of parameters that suggest the highest correlation between \el ~magnitude and its previous calendar year's SysSampEn, during the period between 1984 and 2017.}
\end{table*}

\section*{Discussion}
We have defined the SysSampEn for complex systems and used it to estimate the spatio-temporal disorder degree of temperature variations in the Ni\~{n}o 3.4 region. 
We find that a low degree of horizontal synchronization and a high degree of random temporal variations of SST or near surface air temperature are precursors of a strong \el. Reliable hindcasts and forecasts of \el ~onsets and magnitudes are achieved for \el ~events that occurred during the last 35 years. Our results reveal a high predictability of both the \el ~onsets and magnitudes already before the boreal spring of the \el ~onset year. For the ongoing El Nino, which started in 2018, our method predicts a weak \el ~with a magnitude of $1.11\pm 0.23 ^\circ C$, based only on data until the calendar year 2017. In addition, from Fig. S2-S5 c, we find that for shorter $l_{eff}$ close to half a year, the correlations between \el ~magnitudes and SysSampEn are still high for certain ranges of parameters. This indicates the possibility for even earlier prediction of \el ~magnitudes, however, with lower prediction skill. This question is left for further studies. 

Discussing possible mechanisms related to our findings may help to understand better or even overcome the SPB also in other forecasting models. 
Here we find some clues from the relationship between near surface ocean turbulence and SST variations\cite{Thorpe2007}. Recently, it was discovered that strong \el s are related to intense ocean turbulence, which is characterized by large lateral diffusivity \cite{Schiermeier2015, Russell2017,Busecke2019}. Enhanced lateral diffusivity during \el ~leads to weaker horizontal temperature gradients and higher horizontal mixing results in lower SysSampEn in the Ni\~{n}o 3.4 region. Our further analyses support our conjecture, as shown in Fig. S6, we find that SysSampEn is inversely proportional to the magnitude of \el ~during \el ~periods. 
Memory effects have been reported in many natural systems, such as the climate system \cite{Bunde1998}, physiology \cite{Peng1993, Bunde2000}, and even in seismic activity \cite{Livina2005, Fan2019}.
Remarkably, we also observe that there exist memory effects in the dynamical evolution of the SysSampEn, i.e., a smaller SysSampEn is more likely to be followed by a larger one and a larger follows a smaller one. Fig. S7 demonstrates the SysSampEn of the previous calendar year versus the average SysSampEn of \el ~calendar years (onset calendar year to withdraw calendar year). 
We argue that during the previous calendar year of a strong \el, the near surface lateral diffusivity might be weak, which could be one of the reasons for high SysSampEn in the Ni\~{n}o 3.4 region. However, we think the above hypothesis still needs further analyses based on climate models and observation data. We note that the interannual variability of mesoscale turbulence at the ocean surface has just been found to be regionally correlated with the ENSO indices \cite{Busecke2019}, which supports our hypothesis. It might be used to explain the relationships among SysSampEn, ocean turbulence, and the \el ~magnitude. Furthermore, we also suspect that the high SysSampEn during the previous calendar year of a strong El Niño is related to the storing of more energy for the event.

The theoretical framework developed here has the potential to improve the \el ~forecasting capability with long lead-time and could also be extended to study and improve our knowledge in other complex systems.


\section*{Data}
The ERA-Interim archive at ECMWF (\url{https://apps.ecmwf.int/datasets/}). ERA-Interim (ERA-Interim) is a global atmospheric reanalysis starting from 1979, and is regularly updated. 
In the present work, we used the zero o'clock daily near surface (1000hPa) temperature, downloaded with a spatial (zonal and meridional) resolution of $2.5^\circ \times 2.5^\circ$.  Data for years from 1979 to 2017 were downloaded on October 4, 2018, and data for the last year 2018 was updated on January 29, 2019.

The ERA5 (\url{https://climate.copernicus.eu/climate-reanalysis}) is a climate reanalysis dataset, developed through the Copernicus Climate Change Service (C3S). It is currently available for the period since 1979 within 3 months of real time. The analysis field of ERA5 has a higher spatial resolution of $31$ km and a higher temporal resolution of $1$ hour, compared to ERA-Interim. Data used in the present work is the zero o'clock daily near surface(1000hPa) temperature, download on January 25, 2019, and SST downloaded on January 30, 2019, downloaded with a spatial (zonal and meridional) resolution of $2.5^\circ \times 2.5^\circ$.

The JRA55-do (\url{https://esgf-node.llnl.gov/search/input4mips/}) extends from 1958 to 2018 and is expected to update annually (around April each year). The SST field has a spatial resolution of $1^\circ \times 1^\circ$ and a temporal resolution of $1$ day. Data used in the present work is the daily mean SST, downloaded on November 8, 2018, downloaded with a spatial (zonal and meridional) resolution of $1^\circ \times 1^\circ$.
\section*{Method}
\subsection*{Data Preprocessing}

For each calendar year $y$ since 1984 (the first five years 1979-1983 of the datasets ERA-Interim, ERA5 and ERA5 SST, are used to calculate the first anomaly value for 1984), at each grid point $\alpha$ in the Ni\~{n}o 3.4 region, we calculate the anomalies by substracting the climatological average from the actual temperature and then dividing by the climatological standard deviation. We do this for each calendar day $t$. For simplicity, leap days were excluded. The calculations of the climatological average and standard deviation are based only on the past data up to the year $y$.

\subsection*{System Sample Entropy}
We first define the System Sample Entropy for an arbitrary system. Let's assume we have $N$ interdependent time series $x_\alpha(t)\ (\alpha=1,2,...,N)$ of length $l$ composing the system. 
\begin{enumerate}
\item From each time series, we select sub-records $k$ of length $m < l$, starting at each $q$-th data point, i.e.,
starting at $t= k\times q+1 = 0\times q + 1, 1\times q+1, 2\times q+1,...$, as long as $k\times q+m <= l$.
Thus a specific sub-records is $X_\alpha^k(m,q)=\{x_\alpha(k\times q+1),x_\alpha(k\times q+2),...,x_\alpha(k\times q+m)\}$. Then we select $n$ sub-records from each time series and construct a set of $N\times n$ template vectors from the system, i.e., $\Theta(m,q,n)=\{X_\alpha^k(m,q):0\leq k \leq n-1, 1\leq \alpha \leq N\}$. We assume that two vectors are close (similar) if their Euclidean distance $d(X_\alpha^i(m,q),X_\beta^j(m,q))<\gamma\times max\{\sigma_\alpha,\sigma_\beta\}$ (if $\alpha=\beta$, then $i\neq j$), where $\sigma_\alpha$ and $\sigma_\beta$ are the standard deviations of time series $x_\alpha(t)$ and $x_\beta(t)$ respectively. $\gamma$ defines the similarity criterion and is a nonzero constant.

\item To examine the probability that two time series which are close at $m$ data points still will be close at the next $p$ data points, we construct analogously another set $\Theta(m+p,q,n)$ by selecting sub-records of length $m+p$. To make the number of template vectors of length $m$ equal to that of length $m+p$, we choose $n\leq \frac{l-m-p}{q}+1$. In order to reduce the parameter degrees of freedom and save calculation time, we take $p=q$, then $n\leq \frac{l-m}{p}$. We assume that two template vectors from the set $\Theta(m+p,q,n)$ are close if $d(X_\alpha^i(m+p,q),X_\beta^j(m+p,q))<\gamma\times max\{\sigma_\alpha,\sigma_\beta\}$ (if $\alpha=\beta$, then $i\neq j$).

\item The SysSampEn of the system is defined as $
SysSampEn(m,p,l_{eff}(n),\gamma)=-log(\frac{A}{B})$,
where $A$ is the number of close vector pairs from the set $\Theta(m+p,q,n)$, $B$ is the number of close vector pairs from the set $\Theta(m,q,n)$, and $l_{eff}(n)=n*p+m$, is the number of days since Jan. 1 of each calendar year, used in the calculation of SysSampEn.. 
\end{enumerate}

\subsection*{Parameter Determination for the SysSampEn}
Here we demonstrate how to determine the $SysSampEn(m,p,l_{eff},\gamma)$ parameters for our Ni\~{n}o 3.4 climate system by using the ERA-Interim data. For each calendar year, we define a system composed of $N=22$ (red circles in Fig. \ref{Fig1}) temperature anomalies time series $T_\alpha(t)\ (1\leq \alpha\leq N)$ of length $l=365$ days.

\begin{enumerate}
	\item  We choose the vector lengths $m$ to be $30$ days or $60$ days, and the length increases $p$ to be $15$ days or $30$ days.
	 We focus on a (bi)monthly timescale since El Nino is an interannual phenomenon.
		\item 
		The purpose of the SysSampEn is to quantify the spatial and temporal disorder of a given system. 
		This entails, that if we have a spatially and temporally correlated complex system, represented by time series, and add random terms, e.g, white noise, to each time series, then the SysSampEn of the new system should be, with high probability, larger than the original SysSampEn. Similarly, if we replace the times series in a spatially highly correlated system, with unrelated time series, the SysSampEn should increase. We use these properties as the basis of two tests to determine, for given $m$ and $p$, the values of $l_{eff}$ and $\gamma$, which enable a reliable discrimination between more and less ordered systems. For simplicity, we assume $\gamma$ to be an integer. 
			 
		\begin{enumerate}
			\item Spatial asynchrony test: For a randomly selected year, we choose randomly three neighboring points on the globe $T_{\beta}$, $T_{\beta+1}$ and $T_{\beta+2}$. To construct a highly coupled system, we randomly choose $N=22$ times one of these three nodes. Thus we obtain a system $G_1$ with 22 nodes, where $T_{\beta}$, $T_{\beta+1}$ and $T_{\beta+2}$ might be present in the system with different frequencies. To contrast, we choose randomly 22 unrelated nodes from the globe to create a system $G_2$. We perform this procedure $M$ times. The accuracy is defined as, 
			\begin{equation}
				accuracy = \frac{1}{M}\sum\limits_{i=1}^{M}S_{i}, 
			\label{eq2}
			\end{equation}
			where, 
\begin{numcases}{S_i=}
   1,  & for $SysSampEn_{G_2} > SysSampEn_{G_1}$; \nonumber\\
   0, & for otherwise.
\end{numcases}

In the present study, we used $M = 100$. The accuracy is shown as a function of $l_{eff}$ and $\gamma$ for $m=60,\ p=15$ in Fig. S4 a.

			\item Temporal disorder test: We compare the SysSampEn of an undisturbed climate system $G_1$, here our Ni\~{n}o 3.4 system, with a new system $G_2$, where random numbers have been added to the original time series. 
			The new system is composed of $N=22$ time series $\tilde{T}_{\alpha}(t)=T_\alpha(t)+ R_\alpha(t)$. The $R_\alpha(t)$ are uncorrelated sequences of independent and uniform random numbers in the range $[-0.5*\sigma,0.5*\sigma]$. Here, $\sigma$ is the average of the $N=22$ individual time series' standard deviations between Jan. 1, 1984 and Dec.31, 2018. We perform this procedure $M$ times. The accuracy is defined as in Eq.~\ref{eq2} and is shown as a function of $l_{eff}$ and $\gamma$ for $m=60,\ q=15$ in Fig. S4 b.
		\end{enumerate} 
	
	\end{enumerate}

\subsection*{Forecasting algorithm for \el ~onsets}
We forecast the onset of an \el ~event in the following year if the forecasting index (average forecast over the four datasets) is $\geq0.5^\circ C$ and the observed ONI in December is below $0.5^\circ C$. Otherwise, we forecast the absence of an \el ~onset. The forecasting index is shown in Fig. \ref{Fig3}a as the heights of rectangles.

Please note that the forecasting index used in the present work is calculated based on the significant linear relationship between SysSampEn and the magnitudes of \el ~events that occurred in the period 1984-2017. To forecast the occurrence or absence of \el ~onsets after 2018, one should keep updating the forecasting index once a new \el ~has terminated, by choosing the function of the SysSampEn which gives the highest correlation $r$ with the magnitudes of all terminated \el ~events.

\subsection*{Forecasting algorithm for \el ~magnitudes}
To forecast the magnitude of an \el ~event starting in the year $y$,
\begin{enumerate}[label=(\roman*)]
	\item for one dataset, we determine the parameters of SysSampEn by using the ones that give the highest correlation $r$ with the magnitudes of the \el ~events that occurred before the forecasted event $y$. We regard only parameter combinations which can provide a high accuracy level.
	\item for one dataset, we calculate the best fitting line $Y=a*X+b$ between the \el ~magnitude and the previous calendar year's SysSampEn, by using least square regression. Here $Y$ stands the magnitudes of the \el ~events, and $X$ stands the corresponding previous year's SysSampEn.
Only past events of the forecasted event $y$ are used in the calculation of the best fitting line.

	\item We calculate the SysSampEn in the year $y-1$, and substitute it into the function of the best fitting line. Then we obtain the expected magnitude of \el ~event starting in the calendar year $y$.
	\item Repeat step 1 and 2 for the other datasets. The forecasted magnitude (blue dots in Fig. \ref{Fig3}c) is obtained by taking the average of the four expected magnitudes (``plus'' symbols in Fig. \ref{Fig3}c).
	\item To determine the error bar of our forecasting, we perform the following leave-one-out hindcasts for each of the past events of the forecasted \el ~event $y$:
\begin{enumerate}
	\item the same as (i).
	\item To obtain the leave-one-out hindcasted magnitude of each past event $\bar{y}<y$, we use all events occurred before $y$ except for the hindcasted one to calculate the best fitting line.
	\item We calculate the SysSampEn in the year $\bar{y}-1$, and substitute it into the function of the best fitting line. Then we obtain the expected magnitude of the \el ~event starting in the calendar year $\bar{y}$.
	\item Repeat step 1 and 2 for the other datasets. The leave-one-out hindcasted (orange dots in Fig. \ref{Fig3}b) is obtained by taking the average of the four expected magnitudes (``plus'' symbols in Fig. \ref{Fig3}a).
\end{enumerate}
\end{enumerate}


To forecast the magnitude of the 2018 \el ~event, we substitute for each dataset the SysSampEn value for the year 2017 into the corresponding best fitting linear function, which is determined by all the past \el ~events (except for the 2009 event). Thus we have four individual forecasts, which we average to obtain our final forecast.

\section*{Data availability}
The authors declare that all data which supports the findings are provided with the paper. All data is available from public. Code is available from the corresponding authors upon reasonable request.

\section*{Acknowledgements }

We acknowledge M. J. McPhaden, S. Havlin, Y. Ashkenazy and N. Marwan for their helpful suggestions.
We thank the “East Africa Peru India Climate Capacities — EPICC” project, which is part of the
International Climate Initiative (IKI). The Federal Ministry for the Environment, Nature Conser-
vation and Nuclear Safety (BMU) supports this initiative on the basis of a decision adopted by the
German Bundestag. The Potsdam Institute for Climate Impact Research (PIK) is leading the execution of the project together with its project partners The Energy and Resources Institute (TERI) and the Deutscher Wetterdienst (DWD).
\section*{Author contributions statement}
All authors designed the research, analyzed data, discussed results, and contributed to writing the manuscript.

\section*{Additional information}
Reprints and permissions information is available at www.nature.com/reprints. Correspondence and requests for
materials should be addressed to J.F (jingfang@pik-potsdam.de) or H.J.S (john@pik-potsdam.de).

\section*{Competing financial interests}

The authors declare no competing financial interests.

\end{document}